\documentclass[11pt]{article}
\begin{document}
\title{Logarithmic Conformal Field Theory Through Nilpotent Conformal Dimensions}
\author{S. Moghimi-Araghi \footnote{e-mail: samanimi@rose.ipm.ac.ir} ,
S. Rouhani \footnote{e-mail: rouhani@karun.ipm.ac.ir} and M.
Saadat \footnote{e-mail: saadat@mehr.sharif.ac.ir}\\
\\
Department of Physics, Sharif University of Technology,\\ Tehran,
P.O.Box: 11365-9161, Iran\\ Institute for studies in Theoretical
physics and Mathematics,\\ Tehran, P.O.Box: 19395-5531, Iran}
\maketitle
\begin{abstract}
We study logarithmic conformal field theories (LCFTs) through the
introduction of nilpotent conformal weights. Using this device, we
derive the properties of LCFT's such as the transformation laws,
singular vectors and the structure of correlation functions. We
discuss the emergence of an extra energy momentum tensor, which
is the logarithmic partner of the energy momentum tensor.
\vspace{5mm}\\
{\it PACS}: 11.25.Hf \\
{\it Keywords}: Conformal field theory
\end{abstract}
\section{Introduction}
Since the paper by Belavin, Polyakov and Zamolodchikov \cite{BPZ}
on the determining role of conformal invariance on the structure
of two dimensional quantum field theories, an enormous amount of
work has been done on the role of conformal field theories (CFTs)
in various aspects of physics such as string theory, critical
phenomena and condensed matter physics. Recently Gurarie
\cite{Gur} has pointed to the existence of LCFTs. Correlation
functions in an LCFT may have logarithmic as well as power
dependence. Such logarithmic terms were ruled out earlier due to
requirements such as unitarity or non existence of null states.

In an LCFT, degenerate operators exist which form a Jordan cell
under conformal transformations. In the simplest case one has a
pair $\phi$ and $\psi$ transforming as:

\begin{eqnarray}
\phi(\lambda z)&=&\lambda^{-\Delta}\phi(z) ,\nonumber\\
\psi(\lambda z)&=&\lambda^{-\Delta}[\psi(z)-\phi(z)\ln \lambda] .
\end{eqnarray}
To our knowledge the first time that logarithmic terms appear was
in \cite{RS}. The emergence of the logarithmic term leads to a
host of unusual properties. For instance the two point correlation
functions can be derived from requirements of invariance under
the action of Virasoro algebra generators $L_{0},L_{\pm1}\:$
\cite{CKT}.
\begin{eqnarray}
\left<\phi(z^{\prime})\phi(z)\right>&=&0 ,\nonumber\\
\left<\psi(z^{\prime})\psi(z)\right>&=
&\frac{1}{(z^{\prime}-z)^{2\Delta}}(-2a\ln(z^{\prime}-z)+b) ,\nonumber\\
\left<\psi(z^{\prime})\phi(z)\right>&=&\frac{a}{(z^{\prime}-z)^{2\Delta}}
.
\end{eqnarray}

Generalization to higher dimensions and Jordan cells of a larger
number of degenerate fields were immediate to follow
\cite{GK,RAK}. The development of LCFTs has had two concerns, its
applications and structural matters. A good survey of both aspects
can be found in \cite{CAR,RAH}. On the structural front, the
development has been patchy. Where exactly do LCFTs fit into the
larger family of CFTs requires elucidation. Interesting work on
this issues has been done by Flohr \cite{Fl-CP1} showing that in
the minimal series, if $p$ and $q$ take values which are not
coprime and fall out of the allowed range, LCFTs result. Also,
singular vectors in LCFTs were analyzed by Flohr \cite{Fl-Sing},
using grassman variables, which suggests that there is a connection with
supersymmetry. Connection with supersymmetry has been discussed
elsewhere too \cite{MS}. More recently in papers by Gurarie and
Ludwig \cite{GL} and by Kausch \cite{Kausch} a more direct
connection with supersymmetry has been pointed out. Another
exciting development is the connection between AdS theories and
LCFT \cite{Ikogan}. Indeed, if LCFTs are the natural framework for
the AdS/CFT correspondence \cite{mald}, then the natural
supersymmetry of this theory also suggests that there is a
possible connection between supersymmetry and LCFT. In this paper
we follow up Flohr's \cite{Fl-Sing} idea by considering a nilpotent variable
$\theta$, which is not anticommutative:

\begin{eqnarray}
\theta_{i}^{2}&=&0 ,\nonumber\\
\theta_{i}\theta_{j}&=&\theta_{j}\theta_{i} .
\end{eqnarray}
This is perhaps equivalent to restricting ourselves to the bosonic
section of a supersymmetric theory. We derive the logarithmic
conformal transformations by assuming that conformal dimensions
have a nilpotent part. In fact let us postulate that a primary
field $\Phi(z,\theta)$ exists with the following transformation law
under scaling:

\begin{eqnarray}
\Phi(\lambda z,\theta)&=&\lambda^{-(\Delta+\theta)}\Phi(z,\theta
) .
\end{eqnarray}

However the nilpotency of $\theta$ allows us to expand both sides
of equation (4). Writing $\Phi(z,\theta)=\phi(z)+\theta \psi(z)$
and $\lambda ^{-(\Delta+\theta)} =\lambda^{-\Delta}(1-\theta
$ln$\lambda)$ equation (4) reduces to the transformation laws
given in equation (1). This gives a Jordan cell of rank 2. In
order to get a higher order cell one has to assume a higher level
of nilpotency for $\theta$. In general if the $n$th power of
$\theta$ vanishes we have the following expansion for a field
$\Phi(z,\theta)$

\begin{eqnarray}
\Phi(z,\theta)&=&
\phi_{0}(z)+\phi_{1}(z)\theta+\phi_{2}(z)\theta^{2}+\ldots+\phi_{n-1}(z)\theta^{n-1}
.
\end{eqnarray}

The components ($\phi_{0}\ldots\phi_{n-1}$) constitute a Jordan
cell of rank $n$. To see this it is enough to expand both sides of
equation (4), using equation (5). Any how, as all the known LCFTs
only contain Jordan cells of rank two, we restrict our derivations
to LCFTs of this type. So, through out this paper we shall assume
$\theta_{i}^{2}=0$ . Generalization to higher rank Jordan cells is
straight forward.

This paper is organized as follows. In section
two we derive two and three point correlation functions. Before
deriving four point functions it is necessary to discuss singular
vectors (section 3) in the light of this new construction. In
this section character formula are also derived. In section four
we discuss the structure of the Kac determinant. In section five
we derive four point functions. This leads to generalization of
the hypergeometric function. The details of this calculation is
discussed in appendix B. In section six we discuss the emergence
of an extra energy momentum operator which is partner to the
original energy momentum tensor. It appears that a pseudo
identity operator and a second energy momentum tensor arise in
every LCFT \cite{Gur,GL}. We close this paper by discussing
modifications which arise near a boundary.
\section{Two and three point Correlation functions}
Using transformation law of equation (4) we attempt to derive the
form of correlation functions. Consider the two point functions

\begin{eqnarray}
G(z_{1},z_{2},\theta_{1},\theta_{2})&=&
\left<\Phi_{1}(z_{1},\theta_{1})\Phi_{2}(z_{2},\theta_{2})\right>
.
\end{eqnarray}
Within this correlation function, there exist the correlators
$\langle \phi_{1}(z_{1}) \phi_{2}(z_{2})\rangle$, $\langle
\psi_{1}(z_{1}) \phi_{2}(z_{2})\rangle$, $\langle \phi_{1}(z_{1})
\psi_{2}(z_{2})\rangle$ and $\langle \psi_{1}(z_{1})
\psi_{2}(z_{2})\rangle$, each of them being coefficient of 1,
$\theta_{1}, \theta_{2}$ and $\theta_{1} \theta_{2}$
respectively. So deriving the correlation function (6) and
expanding it in powers of $\theta_{1}$ and $\theta_{2}$, one
easily finds the two point functions of the fields in the theory.

Translation invariance requires $G$, to be a function
of $z_{1}-z_{2}$ only. For rotation and scale invariance we must have
\begin{eqnarray}
G(\lambda(z_{1}-z_{2}),\theta_{1},\theta_{2})&=&
\lambda^{-(\Delta_{1}+\theta_{1})}\lambda^{-(\Delta_{2}+
\theta_{2})}G((z_{1}-z_{2}),\theta_{1},\theta_{2}) ,
\end{eqnarray}
Choosing $\lambda=(z_{1}-z_{2})^{-1}$, we get
\begin{eqnarray}
G(z_{1}-z_{2},\theta_{1},\theta_{2})&=&
\frac{1}{(z_{1}-z_{2})^{\Delta_{1}+\Delta_{2}+
\theta_{1}+\theta_{2}}}f(\theta_{1},\theta_{2}) .
\end{eqnarray}
For symmetry under special conformal transformations

\begin{eqnarray}
z&\longrightarrow&\frac{1}{z+b} ,
\end{eqnarray}
firstly we observe that equation (8) is inconsistent unless
$\Delta_{1}=\Delta_{2}$ or $f=0$. Setting $\Delta_{1}=\Delta_{2}$
we then get:
\begin{eqnarray}
f(\theta_{1},\theta_{2})&=&f(\theta_{1},\theta_{2})
{(z_{1}+b)}^{(\theta_{1}-\theta_{2})}{(z_{2}+b)}^{(\theta_{2}-\theta_{1})}
.
\end{eqnarray}
This does not lead to a vanishing $f$, since $\theta_{i}$'s are
nilpotent. Specializing to the case of a rank 2 Jordan cell, we
have as the most general possible solution of equation (10):
\begin{eqnarray}
f(\theta_{1},\theta_{2})&=&a_{1}(\theta_{1}+\theta_{2})+a_{12}\theta_{1}\theta_{2}
.
\end{eqnarray}
This leads to the two point function

\begin{eqnarray}
\left<\Phi(z_{1},\theta_{1})\Phi(z_{2},\theta_{2})\right>&=&
\frac{1} {(z_{1}-z_{2})^{2\Delta+(\theta_{1}+\theta_{2})}}
(a_{1}(\theta_{1}+\theta_{2})+a_{12}\theta_{1}\theta_{2})
\end{eqnarray}
which is identical to equation (2), if we expand
$\Phi(z_{i},\theta_{i})$. The above results are easily generalized
to a Jordan cell of rank $n$, and also to arbitrary
dimensions.

Let us next consider the three point function

\begin{eqnarray}
G(z_{1},z_{2},z_{3},\theta_{1},\theta_{2},\theta_{3})&=&\left<
\Phi_{1}(z_{1},\theta_{1})\Phi_{2}(z_{2},\theta_{2})
\Phi_{3}(z_{3},\theta_{3})\right>.
\end{eqnarray}
Again it is clear that if one obtains this correlation function,
all the correlators such as
$\langle\phi_{1}\phi_{2}\phi_{3}\rangle$,
$\langle\phi_{1}\phi_{2}\psi_{3}\rangle$, $\cdots$ can be
calculated readily by expanding this correlation function in
terms of $\theta_{1}$, $\theta_{2}$ and $\theta_{3}$. Also note
that these fields may belong to different Jordan cells. The
procedure of finding this correlator is just the same as the one
we did for the two point function. Like ordinary CFTs, the three
point function is obtained up to some constants. Of course, in
our case it is found up to a function of $\theta_{i}$'s, {\it
i.e.}

\begin{eqnarray}
G(z_{1},z_{2},z_{3},\theta_{1},\theta_{2},\theta_{3})&=&
f(\theta_{1},\theta_{2},\theta_{3})z_{12}^{-a}z_{23}^{-b}z_{31}^{-c},
\end{eqnarray}
where $z_{ij}=(z_{i}-z_{j})$ and
\begin{eqnarray}
a&=&\Delta_{1}+\Delta_{2}-\Delta_{3}+(\theta_{1}+\theta_{2}-\theta_{3}) ,\nonumber\\
b&=&\Delta_{2}+\Delta_{3}-\Delta_{1}+(\theta_{2}+\theta_{3}-\theta_{1}) ,\nonumber\\
c&=&\Delta_{3}+\Delta_{1}-\Delta_{2}+(\theta_{3}+\theta_{1}-\theta_{2})
.
\end{eqnarray}
There are some constraints on
$f(\theta_{1},\theta_{2},\theta_{3})$, but further reduction
requires specification the rank of Jordan cell. As we have taken
it to be 2, we have:
\begin{eqnarray}
f(\theta_{1},\theta_{2},\theta_{3})=\sum_{i\neq j\neq
k}^{3}C_{i}(\theta_{j}+\theta_{k}) +\sum_{1\leq
i<j\leq3}C_{ij}\theta_{i}\theta_{j}
+C_{123}\theta_{1}\theta_{2}\theta_{3} .
\end{eqnarray}
While symmetry considerations do not rule out a constant term on
the right hand side of equation (16), but a consistent OPE forces
this constant to vanish \cite{MRS}. This form together with the
equations (14) - (16) leads to correlation functions already
obtained in the literature. This is also consistent with the
observation that in all the known LCFT's so far the three point
function of the first field in the Jordan cell vanishes.

In a similar fashion one can derive the form of the four point
functions. But before this is done, we need to address the
question of singular vectors in an LCFT.
\section{Singular Vectors in LCFT}
Considering the infinitesimal transformation consistent with
equation (4) we have :
\begin{eqnarray}
\delta\Phi&=&-\varepsilon (z^{n+1}\frac{\partial}{\partial
z}+(n+1)(\Delta+\theta)z^{n})\Phi .
\end{eqnarray}
This defines the action of the generators of the Virasoro algebra
on the primary fields and points to the existence of a highest
weight vector with nilpotent eigenvalue:
\begin{eqnarray}
L_{0}|\Delta+\theta\rangle&=&(\Delta+\theta)|\Delta+\theta\rangle ,\nonumber\\
L_{n}|\Delta+\theta\rangle&=&0,\:\:\:\:\:\:\:\:\:n\geq1\:\:.
\end{eqnarray}
Nilpotent state $|\Delta+\theta\rangle$ can be considered as:
\begin{equation}
|\Delta+\theta\rangle=|\phi \rangle+ \theta | \psi \rangle .
\end{equation}
It can be easily seen that the law written in equation (19), leads
to the well known equations:
\begin{eqnarray}
L_{0}|\phi\rangle=\Delta|\phi\rangle ,\nonumber\\
L_{0}|\psi\rangle=\Delta|\psi\rangle+|\phi\rangle .
\end{eqnarray}
However the norm of this vector is complex. As discussed in
\cite{Kogan}, the norm of the state $|\phi\rangle$ is zero and
the states $|\phi\rangle$ and $|\psi\rangle$ are not orthogonal to
each other. Instead of orthogonality condition, one can choose
\begin{equation}
\langle\psi|\phi\rangle = \langle\phi|\psi\rangle =1 .
\end{equation}
The norm of $|\psi\rangle$ is not well defined and it is taken to
be $ d $. Putting all these together, one can define:
\begin{eqnarray}
\langle\Delta+\theta|\Delta+\theta\rangle=\theta+
\bar{\theta}+d\:\: \bar{\theta}\theta .
\end{eqnarray}
In addition to these highest weight states, there are descendants
which can be obtained by applying $L_{-n}$'s on the highest
weight vectors:
\begin{eqnarray}
|\Delta+n_{1}+n_{2}+\cdots +n_{k}+\theta\rangle &=&
L_{-n_{1}}L_{-n{2}}\cdots L_{-n_{k}}|\Delta+\theta\rangle.
\end{eqnarray}
To acquire more information about the content of descendants
level by level, and hence the secondary operators, one usually
computes the character formula:
\begin{eqnarray}
\chi_{\Delta}(\theta,\bar{\theta})&=&\sum_{N}\langle
N+\Delta+\theta| \eta^{L_{0}-\frac{c}{24}}|N+\Delta+\theta\rangle
\end{eqnarray}
which by equation (23) simplifies to
\begin{eqnarray}
\chi_{\Delta}(\theta,\bar{\theta})&=&\eta^{\Delta+\theta-\frac{c}{24}}
\sum_{N} \eta^{N}g(N,\theta)\langle \Delta+\theta
|\Delta+\theta\rangle .
\end{eqnarray}
Writing $g(N,\theta)=g_{0}(N)+\theta g_{1}(N)$ we obtain four
characters:
\begin{eqnarray}
\chi_{\Delta}^{(\phi,\phi)}&=&0 ,\nonumber\\
\chi_{\Delta}^{(\phi,\psi)}&=&\chi_{\Delta}^{(\psi,\phi)}=
\eta^{\Delta-\frac{c}{24}}\sum_{N} \eta^{N}g_{0}(N) ,\nonumber\\
 \chi_{\Delta}^{(\psi,\psi)}&=&\eta^{\Delta-\frac{c}{24}}\sum_{N}
\eta^{N}\left[g_{1}(N)+(d+\ln\eta)g_{0}(N)\right] .
\end{eqnarray}
Appearance of logarithms in character formula have been discussed
in \cite{Fl-CP1,Kogan}.

There is a submodule in which states transform among themselves,
under any conformal transformation. Such a submodule is generated
from a state of the form given in equation (19), such that
$L_{k}|\chi_{\Delta,c}^{n}(\theta)\rangle
=0\:\:\:\:,\:\:\:k\geq1$. Properties of $L_{k}$ imply that it is
sufficient to have:
\begin{eqnarray}
L_{k}|\chi_{\Delta,c}^{n}(\theta)\rangle&=&0 ,
\:\:\:\:\:\:\:\:\:\:\:k=1,2 .
\end{eqnarray}
Such a vector is a linear combination of descendant vectors of
level $n$, so we can write:
\begin{eqnarray}
|\chi_{\Delta,c}^{n}(\theta)\rangle&=&
\sum_{\{n_{1}+n_{2}+\ldots+n_{m}=n\}} b^{(n_{1},n_{2},\cdots,
n_{m})}L_{-n_{m}}\ldots L_{-n_{1}}|\Delta+\theta\rangle .
\end{eqnarray}
 A singular vector must also be orthogonal to the whole vector
module and in particular itself. At level $n$, there are $p(n)$,
(partition of n) unknown coefficients in a singular vector
$|\chi_{\Delta,c}^{n}\rangle$ which need to be determined. The
action of $L_{1}$ on equation (28) provides $p(n-1)$ relations
among the coefficients. One of the coefficients is arbitrary and the requirement that
$L_{2}|\chi_{\Delta,c}^{n}(\theta)\rangle=0$, gives $p(n-2)$
relations. In the following we determine the null vectors at level
2 for a rank 2 Jordan cell. We thus have:
\begin{eqnarray}
|\chi_{\Delta,c}^{2}(\theta)\rangle&=&\left(b^{(1,1)}L_{-1}^{2}+b^{(2)}L_{-2}\right)|\Delta+\theta\rangle
,
\end{eqnarray}
by applying $L_{1}$ we have:
\begin{eqnarray}
\left[b^{(1,1)}[4(\Delta+\theta)+2]L_{-1}+3b^{(2)}L_{-1}\right]|\Delta+\theta\rangle&=&0
.
\end{eqnarray}
Thus we have the solution $b^{(1,1)}=3$ and
\begin{eqnarray}
b^{(2)}&=&-[4(\Delta+\theta)+2] .
\end{eqnarray}
Further more
\begin{eqnarray}
[6b^{(1,1)}(\Delta+\theta)+b^{(2)}(4\Delta+\frac{c}{2}+4\theta)]|\Delta+\theta\rangle&=&0
.
\end{eqnarray}
We thus get:
\begin{eqnarray}
16\Delta^{2}+2\Delta(c-5)+c+2\theta(16\Delta+c-5)&=&0 .
\end{eqnarray}
We thus observe that for $\Delta=-\frac{5}{4}$ or $\frac{1}{4}$, a
logarithmic null vector exists if $c=25$ and 1 respectively.
Therefore the if we write
$|\chi_{\Delta,c}^{2}(\theta)\rangle=|\chi_{\Delta,c}^{2}(0)\rangle+
\theta|\chi_{\Delta,c}^{2}(1)\rangle$ then
\begin{eqnarray}
|\chi_{\Delta,c}^{2}(0)\rangle&=&\left[3L_{-1}^{2}-(4\Delta+2)L_{-2}\right]|\phi\rangle ,\nonumber\\
|\chi_{\Delta,c}^{2}(1)\rangle&=&\left[3L_{-1}^{2}-(4\Delta+2)L_{-2}\right]
|\psi\rangle-4L_{-2}|\phi\rangle .
\end{eqnarray}
By the same technique logarithmic singular vectors can be obtained
at higher levels. In the appendix A we present a level 3 singular
vector. These results are consistent with findings of
\cite{Fl-Sing}. However there is evidence that, this method does
not give all singular vectors \cite{Private}, perhaps conditions
of equation (27) are too strong.
\section{Kac determinant in LCFT}
In section 3 we discussed the singular vectors which occur at
level 2 and 3 for a rank 2 Jordan cell. We saw that unlike
ordinary CFT in an LCFT, for some special values of $\Delta$ and
c singular vectors exist. Our technique in section 3 for finding
singular vectors was that, we demanded at level $n$,
$|\chi_{\Delta,c}^{n}(\theta)\rangle$ be a primary field in the
sense that $L_{1}|\chi_{\Delta,c}^{n}(\theta)\rangle=L_{2}
|\chi_{\Delta,c}^{n}(\theta)\rangle=0$. As in an ordinary CFT we
can relate the discussion of singular vectors in LCFT to Kac
determinant. So it is necessary to find the form of the Kac
determinant in the nilpotent $\theta$ formalism.

In ordinary CFTs, the Kac determinant arises out of setting
$\langle \Delta|M|\Delta\rangle$ to zero where the elements of $M$
have a general form like $L_{i}^{m}L_{j}^{n}\cdots$. The nonzero
contributions only come from parts like $L_0^k$ and these terms
are replaced by $\Delta^k\langle \Delta|\Delta\rangle=\Delta^k$.
In our case all the above is true except that $|\Delta\rangle$
should be replaced by $|\Delta + \theta\rangle$, and since
$L_0|\Delta + \theta\rangle = (\Delta + \theta)|\Delta +
\theta\rangle$ all $\Delta$'s are replaced by $\Delta + \theta$.
However now the norm of the state $|\Delta + \theta\rangle$ is
given by equation (22), and all the coefficients of $\theta$ and
$\bar{\theta}$ have to vanish independently. One thus concludes
that the determinant of the matrix $M$ should vanish. So, at
level $n$ Kac determinant has the form:
\begin{equation}
{\det}_{n}(c,\Delta+\theta)=\prod_{r,s=1;1 \leq r s\leq
n}^{n}\left(\Delta+\theta-\Delta_{r,s}(c)\right)^{p(n-rs)} ,
\end{equation}
where $\Delta_{r,s}(c)$ is
\begin{equation}
\Delta_{r,s}(c)=
\frac{1}{96}\left[(r+s)\sqrt{1-c}-(r-s)\sqrt{25-c}\right]^{2}
-\frac{1-c}{24} .
\end{equation}
and $p(n-rs)$ is the number of partitions of the integer $n-rs$.
In an ordinary CFT when the Kac determinant vanishes we have a
singular vector. In our case and for a rank 2 Jordan cell the
condition of vanishing $\det_{n}(c,\Delta+\theta)$ are:

(i) If $p(n-rs)\geq2$ for some $r$ and $s$, Kac determinant
vanishes for all values of $\Delta$ that satisfy in
$\Delta=\Delta_{r,s}(c)$.

(ii) If $p(n-rs)=1$ for some pairs of
$(r,s)=(r_{1},s_{1}),(r_{2},s_{2}),\cdots$ we can have vanishing
determinant if at least
$\Delta=\Delta_{r_{i},s_{i}}(c)=\Delta_{r_{j},s_{j}}(c)$. In this
case unlike (i) we are limited to special values for $\Delta$ and
$c$ which last condition is held. As an example we consider Kac
determinant at level 3. Since
\begin{equation}
p(3-rs)=\cases{2&\mbox r=1 , s=1 \cr
               1&\mbox r=1 , s=2 or r=2 , s=1 \cr
               1&\mbox r=1 , s=3 or r=3 , s=1 } ,
\end{equation}
and $\Delta_{1,1}=0$ and other $\Delta_{r,s}$ are nonzero,
therefore $\Delta_{1,1}$ is a case of (i) and others are cases of
(ii)
\begin{eqnarray}
\Delta_{1,3}&=&\Delta_{3,1}\Rightarrow \cases{c=1 , \Delta=1 \cr
                                   c=25 , \Delta=-3} ,\nonumber\\
\Delta_{1,2}&=&\Delta_{2,1}\Rightarrow \cases{c=1 ,
\Delta=\frac{1}{4} \cr
                                   c=25 ,
                                   \Delta=\frac{-5}{4}} ,\nonumber\\
\Delta_{1,3}&=&\Delta_{2,1}\Rightarrow c=28 ,\Delta=-2 ,\nonumber\\
\Delta_{3,1}&=&\Delta_{1,1}\Rightarrow c=-2 ,\Delta=0 .\\
\end{eqnarray}
These results are consistent with those of \cite{Fl-Sing}. We
observe that LCFTs are possible only when the Kac determinant has
multiple zeros. This means that we have an LCFT only when
degenerate conformal weights exist. Cases have been seen before,
the simplest LCFT is $c=-2$ first given by \cite{Gur}.
\section{Four point functions}
To obtain further information about the theory with which we are
concerned, such as surface critical exponents, OPE structure,
monodromy group etc. one should compute four point correlation
functions. In the language we have developed so far, the four
point correlation functions depend on four $\theta$'s in addition
to the coordinates of points:
\begin{eqnarray}
G(z_{1},z_{2},z_{3},z_{4},\theta_{1},\theta_{2},\theta_{3},
\theta_{4})&=&\left<\Phi_{1}(z_{1},\theta_{1})\ldots\Phi_{4}
(z_{4},\theta_{4})\right>\nonumber\\
&=&f(\eta,\theta_{1},\theta_{2},\theta_{3},\theta_{4})\prod_{1\leq
i\leq j\leq4}z_{ij}^{\mu_{ij}} .
\end{eqnarray}
where
\begin{eqnarray}
\mu_{ij}&=&\frac{1}{3}\sum_{k=1}^{4}(\Delta_{k}+\theta_{k})
-(\Delta_{i}+\theta_{i})-(\Delta_{j}+\theta_{j})
\:\:\:\:\:\:\:\:\:\:,\:\:\:\:\:\:\:\:\:\:\:\:\:
\eta=\frac{z_{41}z_{23}}{z_{43}z_{21}}
\end{eqnarray}

$\:\:\:\:\:\:\:\:\:\:\:\:\:\:\:\:\:\:$

This form is invariant under all conformal transformations.
Although there is no other restrictions on $G$ due to symmetry
considerations, but because of OPE structure, the four-point
function $\langle\phi\phi\phi\phi\rangle$ should vanish
\cite{MRS,FlohrNew}, that is, the term independent of $\theta_i$'s
in $G$ is zero. Thus in addition to the differential equations
which should be satisfied by $G$, one must impose the condition
$\langle\phi\phi\phi\phi\rangle=0$ on the solution derived.

If there is a singular vector in the theory, a differential
equation can be derived for
$f(\eta,\theta_{1},\theta_{2},\theta_{3},\theta_{4})$. Let us
consider a theory which contains a singular vector of level two.
As seen in previous section the singular vector in such a theory
is:
\begin{equation}
\chi^{(2)}(z_{4},\theta_{4})=\left[3
L_{-1}^{2}-(2(2\Delta_{4}+1)+4\theta_{4})L_{-2}\right]\Phi_{4}(z_{4},\theta_{4})
.
\end{equation}
As this vector is orthogonal to all the other operators in the
Verma module
\begin{eqnarray}
\langle\Phi_{1}\Phi_{2}\Phi_{3}\chi^{(2)}\rangle&=&0 ,
\end{eqnarray}
one immediately is led to the differential equation:

\begin{equation}
\left[3\partial_{z_{4}}^{2}-(2(2\Delta_{4}+1)+4\theta_{4})
\sum_{i=1}^{3}\frac{\Delta_{i}+\theta_{i}}{(z_{i}-z_{4})^{2}}
-\frac{\partial_{z_{i}}}{z_{i}-z_{4}}\right]\langle\Phi_{1}
\Phi_{2}\Phi_{3}\Phi_{4}\rangle=0 .
\end{equation}
By sending points to
$z_{1}=0,\:\:\:z_{2}=1,\:\:\:z_{3}\longrightarrow\infty$ and
$z_{4}=\eta, $ we find:
\begin{eqnarray}
\partial_{\eta}^{2}f+[\frac{2\mu_{14}}{\eta}-\frac{2\mu_{24}}
{1-\eta}-\alpha\frac{2\eta-1}{\eta(1-\eta)}]\partial_{\eta}f
+[\frac{\mu_{14}(\mu_{14}-1)}{\eta^{2}}+\frac{\mu_{24}(\mu_{24}-1)}
{(1-\eta)^{2}}\nonumber\\
-\frac{2\mu_{14}\mu_{24}}{\eta(1-\eta)}-\frac{\alpha(\Delta_{1}+
\theta_{1}-\mu_{14})}{\eta^{2}}-\frac{\alpha(\Delta_{2}+\theta_{2}
-\mu_{24})}{(1-\eta)^{2}}+\frac{\alpha\mu_{12}}{\eta(1-\eta)}]f&=&0
,
\end{eqnarray}
where $\alpha=\frac{1}{3}\left[2(2\Delta_{4}+1)
+4\theta_{4}\right]$. Renormalizing using
\begin{eqnarray}
H(\eta,\theta_{1},\theta_{2},\theta_{3},
\theta_{4})&=&\eta^{-\beta_{1}+\mu_{14}}(1-\eta)^{-\beta_{2}+\mu_{24}}f(\eta,\theta_{1},\theta_{2},\theta_{3},
\theta_{4}) ,
\end{eqnarray}
we find that $\beta_{i}$ satisfy
\begin{eqnarray}
\beta_{i}(\beta_{i}-1)+\alpha(\beta_{i}-
\Delta_{i}-\theta_{i})&=&0 ,\:\:\:\:\:\:\:\:\:\:i=1,2
\end{eqnarray}
and $H$ satisfies the hypergeometric equation:
\begin{eqnarray}
\eta(1-\eta)\frac{d^{2}H}{d\eta^{2}}+[c-(a+b+1)\eta]\frac{d
H}{d\eta}- a b H&=&0 .
\end{eqnarray}
where
\begin{eqnarray}
ab&=&(\beta_{1}+\beta_{2})(\beta_{1}+
\beta_{2}+2\alpha-1)+\alpha(\Delta_{4}-
\Delta_{3}+\theta_{4}-\theta_{3}) ,\nonumber\\
a+b+1&=&2(\beta_{1}+\beta_{2}+\alpha) ,\nonumber\\
c&=&2\beta_{1}+\alpha .
\end{eqnarray}
We can now write down the solution of equation (48) in terms of
the hypergeometric series
\begin{equation}
H(a,b,c;\eta)=K(\theta_{1},\theta_{2},\theta_{3},\theta_{4})h(a,b,c;\eta),
\end{equation}
where
\begin{equation}
 h(a,b,c;\eta)=\sum_{n=0}^{\infty}\frac{(a)_{n}(b)_{n}}{n!(c)_{n}}\eta^{n} ,
\end{equation}
with $ (x)_{n}=x(x+1)\ldots(x+n-1)\:\:,(x)_{0}=1\:\:.$ And
\begin{equation}
K(\theta_{1},\theta_{2},\theta_{3},\theta_{4})=
\sum_{i=1}^{4}k_{i}\theta_{i}+\sum_{1\leq
i<j\leq4}k_{ij}\theta_{i}\theta_{j}+ \sum_{1\leq
i<j<k\leq4}k_{ijk}\theta_{i}\theta_{j}\theta_{k}+k_{1234}
\theta_{1}\theta_{2}\theta_{3}\theta_{4}.
\end{equation}

Note that the coefficients in equation (51) contain nilpotent
terms. Therefore equation (51) actually describes more than one
solution. This expansion results in 16 functions. This is natural
because as seen in the case of two and three point functions,
inside the correlator (40) there exist sixteen distinct
correlation functions. Of course, if all the fields belong to the
same Jordan cell, only four of them may be independent and the
rest are related by crossing symmetry. Also, one of them which
only contains $\phi$ fields, vanishes and hence only three
independent functions remain. The form of these functions may be
obtained by expanding equation (50) and collecting powers of
$\theta_{i}$'s. Note that expanding equation (48) leads to
sixteen differential equations all of which are not independent.
The general form of these equations is given in appendix B. As an
example we solve equation (48) for the special case of $
\Delta_{1}=\Delta_{2}=\Delta_{3}=\Delta_{4}=\frac{1}{4}$. In this
case one of the solutions of equation (47) is:
\begin{eqnarray}
\beta_{1}&=&\frac{1}{2}+\theta_{1}-\frac{1}
{3}\theta_{4}+\frac{2}{3}\theta_{1}\theta_{4} ,\nonumber\\
\beta_{2}&=&\frac{1}{2}+\theta_{2}-\frac{1}{3}
\theta_{4}+\frac{2}{3}\theta_{2}\theta_{4} .
\end{eqnarray}
We then find from equation (49):
\begin{eqnarray}
a&=&2+\theta_{1}+\theta_{2}+\theta_{3}+\theta_{4}
+\frac{2}{3}(\theta_{1}+\theta_{2}+\theta_{3}) \theta_{4} ,
\nonumber\\ b&=&1+\theta_{1}+\theta_{2}-\theta_{3}+
\frac{1}{3}\theta_{4}+\frac{2}{3}(\theta_{1}+\theta_{2}-\theta_{3})
\theta_{4} ,\nonumber\\
c&=&2+2\theta_{1}+\frac{2}{3}\theta_{4}+\frac{4}{3}\theta_{1}
\theta_{4} .
\end{eqnarray}
Finally from equations (40), (46) and (50) we get expressions for
the various four point functions. For example
\begin{eqnarray}
\left<\psi(z_{1})\phi(z_{2})\phi(z_{3})\phi(z_{4})
\right>=k_{1}\eta^{\frac{2}{3}}(1-\eta)^{\frac{2}{3}}h_{0}(\eta)\prod_{1\leq
i<j\leq4}z_{ij}^{-\frac{1}{6}} .
\end{eqnarray}
Where $h_{0}(\eta)$ is given in appendix B. The other four point
functions having 2, 3 or 4 $\psi$'s can be calculated in the same
way. In these functions, depending on how many $\psi$'s are
present in the correlators, different functions appear on the
right hand side. If there is no $\psi$ in the correlator, the
correlator is zero, if there is one $\psi(z_{i})$, only $H_{i}$
appears, if there are two $\psi$'s, $H_{i}$ and $H_{ij}$ appear,
and so on. The situation is just the same for the two or three
point functions. For example the two point function $\langle \phi
(z) \psi(0) \rangle$ is written in terms of the function
$z^{-2\Delta}$ and in the correlator $\langle \psi (z) \psi(0)
\rangle$ there exist both functions $z^{-2\Delta}$ and
$z^{-2\Delta}$ln$z$.
\section{Energy momentum tensor}
Two central operators in a CFT are the energy momentum tensor,
$T$ with conformal weight $\Delta=2$ and the identity operator,
$I$ with conformal weight $\Delta=0$. However, $T$ is a secondary
field of the identity, because $L_{-2}I=T$.

 In an LCFT degenerate
operators exist which form a Jordan cell under conformal
transformation. This holds true for the identity as well. The
existence of a logarithmic identity operator has been discussed
by a number of authors \cite{Gur,I1,I2}.

 Now consider the
identity operator $I_{0}$ and its logarithmic partner $I_{1}$.
According to equation (1) this pair transforms as:
\begin{eqnarray}
I_{0}(\lambda z)&=&I_{0}(z) ,\nonumber\\ I_{1}(\lambda
z)&=&I_{1}(z)-I_{0}(z)\ln \lambda .
\end{eqnarray}
So according to our convention, we define a primary field
$I(z,\theta)$:
\begin{equation}
I(z,\theta)=I_{0}(z)+\theta I_{1}(z) .
\end{equation}
with conformal weight $\theta$. Under scaling, $I(z,\theta)$
transforms according to equation (4). Thus we have:
\begin{eqnarray}
L_{0}I_{0}(z)&=&0 ,\nonumber\\ L_{0}I_{1}(z)&=&I_{0}(z).
\end{eqnarray}
this was first observed in $c=-2$ theory by Gurarie \cite{Gur}.

Here we wish to find the field $T(z,\theta)$ with conformal weight
$2+\theta$ which is a secondary of $I(z,\theta)$ in the sense:
\begin{equation}
L_{-2}I(z,\theta)=T(z,\theta) .
\end{equation}
By writing $T(z,\theta)=T_{0}(z)+\theta T_{1}(z)$ and since
$L_{0}T(z,\theta)=(2+\theta)T(z,\theta)$ we have:
\begin{eqnarray}
L_{0}T_{0}(z)&=&2T_{0}(z) ,\nonumber\\
L_{0}T_{1}(z)&=&2T_{1}(z)+T_{0}(z) .
\end{eqnarray}
This points to the existence of an extra energy momentum tensor
\cite{GL,I1,I2}. By applying $L_{2}$ on both sides of equation
(59) we have:
\begin{eqnarray}
L_{2}T_{0}(z)&=&\frac{c}{2}I_{0}(z) ,\nonumber\\
L_{2}T_{1}(z)&=&\frac{c}{2}I_{1}(z)+4I_{0}(z) .
\end{eqnarray}

The first of this pair exists in an ordinary CFT, so $T_{0}(z)$
leads to the Virasoro algebra, while $T_{1}(z)$ must leads to a
new algebra \cite{GL}. We now attempt at finding the OPE of the
extra energy momentum tensor, $T_{1}(z)$ with $\Phi(z,\theta)$ and
$T(z,\theta)$. Because of OPE's invariance under scaling and
according to our convention it is sufficient to change conformal
weight of each field to $\Delta+\theta$. Consider the following OPE:
\begin{equation}
T_{0}(z^{\prime})\Phi(z,\theta)=\frac{\Delta+\theta}{(z^{\prime}-z)^{2}}\Phi(z,\theta)+
\frac{\partial_{z}\Phi(z,\theta)}{z^{\prime}-z}+\cdots\:\:\:.
\end{equation}
This relation leads to the familiar OPE for $T(z^{\prime})\phi(z)$
a new OPE:
\begin{equation}
T_{0}(z^{\prime})\psi(z)=\frac{\phi(z)+\Delta\psi(z)}{(z^{\prime}-z)^{2}}+
\frac{\partial_{z}\psi(z)}{z^{\prime}-z}+\cdots\:\:\:.
\end{equation}
Also
\begin{equation}
T_{0}(z^{\prime})T(z,\theta)=\frac{\frac{c(\theta)}{2}I(\theta)}{(z^{\prime}-z)^{4}}
+\frac{2+\theta}{(z^{\prime}-z)^{2}}T(z,\theta)+\frac{\partial_{z}T(z,\theta)}{z^{\prime}-z}+\cdots\:\:\:,
\end{equation}
where $c(\theta)=c_{1}+\theta c_{2}$. Again we obtain two OPE, one
of them is $T(z^{\prime})T(z)$ which is known from CFT and the
other is:
\begin{equation}
T_{0}(z^{\prime})T_{1}(z)=\frac{\frac{c_1}{2}I_1(z)+\frac{c_2}{2}I_0}{(z^{\prime}-z)^{4}}
+\frac{T(z)+2T_{1}(z)}{(z^{\prime}-z)^{2}}+\frac{\partial_{z}T_{1}(z)}{z^{\prime}-z}+\cdots\:\:\:.
\end{equation}

The emergence of an extra energy momentum tensor and central
charge have been noticed by Gurarie and Ludwig \cite{GL},
although our approach is very different.

It is worth noting that equations (56) imply that $\langle I_0
\rangle$ vanishes whereas $\langle I_1 \rangle = 1$ . This
immediately results in the vanishing of $\langle T_{0}T_{0}
\rangle $.

\section{Boundary}
Let us now consider the problem of LCFT near a boundary. As shown
in \cite{Cardy} in an ordinary CFT, if the real axis is taken to
be the boundary, with certain boundary condition that
$T=\overline{T}$ on the real axis, the differential equation
satisfied by n-point function near a boundary are the same as the
differential equations satisfied by 2n-point function in the
bulk. This trick may be used in order to derive correlations of
an LCFT near a boundary \cite{Kogan,MR}. Here we rederive the same
results using the nilpotent formalism. Again we consider an LCFT
with a rank 2 Jordan cell. First we find the one point functions
of this theory. By applying $L_{0},L_{\pm1}$ on the correlators,
one obtains:
\begin{eqnarray}
(\partial_{z}+\partial_{\bar{z}})\langle\Phi(z,\bar{z},\theta)
\rangle&=&0 ,\nonumber\\
(z\partial_{z}+\bar{z}\partial_{\bar{z}}+2(\Delta+\theta)
\langle\Phi(z,\bar{z},\theta\rangle&=&0 ,\nonumber\\
(z^{2}\partial_{z}+\bar{z}^{2}\partial_{\bar{z}}+
2z(\Delta+\theta)+2\bar{z}(\Delta+\theta))\langle
\Phi(z,\overline{z},\theta\rangle&=&0 .
\end{eqnarray}
In these equations, we have assumed that $\Phi$ is a scalar field
so that $\Delta=\bar{\Delta}$. The first equation states
$\langle\Phi(z,\bar{z},\theta)\rangle$ is a function of
$z-\bar{z}$ and the solution to the second equation is:
\begin{eqnarray}
\langle\Phi(y,\theta)\rangle&=&\frac{f(\theta)}{y^{2(\Delta+\theta)}}
,
\end{eqnarray}
where $y=z-\bar{z}$. The third line of equation (66) is
automatically satisfied by this solution. Expanding $f(\theta)$
as $a+b\theta$ one finds:
\begin{eqnarray}
\langle\Phi(y,\theta)\rangle&=&\frac{a}{y^{2\Delta}}+\frac{\theta}{y^{2\Delta}}(b-2a\ln
y).
\end{eqnarray}
As the field $\Phi(y,\theta)$ is decomposed to
$\phi(y)+\theta\psi(y)$ one can read the one-point functions
$\langle\phi(y)\rangle$ and $\langle\psi(y)\rangle$ from the
equation (68):
\begin{eqnarray}
\langle\phi(y)\rangle&=&\frac{a}{y^{2\Delta}},
\end{eqnarray}
\begin{eqnarray}
\langle\psi(y)\rangle&=&\frac{1}{y^{2\Delta}}(b-2a\ln y) .
\end{eqnarray}
To go further, one can investigate the two-point function
$G(z_{1},\bar{z}_{1},z_{2},\bar{z}_{2},\theta_{1},\theta_{2})=
\langle\Phi(z_{1},\bar{z}_{1},\theta_{1})\Phi(z_{2},
\bar{z}_{2},\theta_{2})\rangle$ in the same theory. Invariance
under the action of $L_{-1}$ implies:
\begin{eqnarray}
(\partial_{z_{1}}+\partial_{\bar{z}_{1}}+\partial_{z_{2}}+\partial_{\bar{z}_{2}})G&=&0
.
\end{eqnarray}
The most general solution of this equation is
$G=G(y_{1},y_{2},x_{1},x_{2},\theta_{1},\theta_{2})$, where
$y_{1}=z_{1}-\bar{z}_{1},\:\:\:\:y_{2}=z_{2}-\bar{z}_{2},
\:\:\:\:x=x_{2}-x_{1}$ and $x_{i}=z_{i}+\bar{z}_{i}$. By
invariance under the action of $L_{_{0}}$ we should have :
\begin{eqnarray}
[y_{1}\frac{\partial}{\partial
y_{1}}+y_{2}\frac{\partial}{\partial
y_{2}}+x\frac{\partial}{\partial
x}+2(\Delta+\theta_{1})+2(\Delta+\theta_{2})]G&=&0 ,
\end{eqnarray}
which implies:
\begin{eqnarray}
G&=&\frac{1}{x^{4\Delta+2\theta_{1}+2\theta_{2}}}
f(\alpha_{1},\alpha_{2},\theta_{1},\theta_{2}) .
\end{eqnarray}
where $\alpha_{1}=\frac{y_{1}}{x}$ and
$\alpha_{2}=\frac{y_{2}}{x}$. Now consider the action of $L_{1}$
on $G$:
\begin{eqnarray}
(x_{1}+x_{2})[y_{1}\frac{\partial}{\partial
y_{1}}+y_{2}\frac{\partial}{\partial
y_{2}}+x\frac{\partial}{\partial
x}+2(\Delta+\theta_{1})+2(\Delta+\theta_{2})]G\nonumber\\
+[xy_{1}\frac{\partial}{\partial
y_{1}}-xy_{2}\frac{\partial}{\partial
y_{2}}+(y_{1}^{2}-y_{2}^{2})\frac{\partial}{\partial x}
+2x(\theta_{1}-\theta_{2})]G=0 .
\end{eqnarray}
The first bracket is zero because of equation (72). Substituting
the solution (73) in equation (74) the function $f$ satisfies:
\begin{eqnarray}
\left(\alpha_{1}+\frac{\alpha_{1}}{\alpha_{1}^{2}-
\alpha_{2}^{2}}\right)\frac{\partial f }{\partial
\alpha_{1}}+\left(\alpha_{2}+\frac{\alpha_{2}}{\alpha_{2}^{2}
-\alpha_{1}^{2}}\right)\frac{\partial f }{\partial
\alpha_{2}}+2\left(2\Delta+\theta_{1}+\theta_{2}+\frac{\theta_{1}
-\theta_{2}}{\alpha_{1}^{2}-\alpha_{2}^{2}}\right)f=0.
\end{eqnarray}
The most general solution of above is:
\begin{eqnarray}
f(\alpha_{1},\alpha_{2})&=&\frac{1}{(\alpha_{1}\alpha_{2})^
{2\Delta+\theta_{1}+\theta_{2}}}
\left(\frac{\alpha_{2}}{\alpha_{1}}\right)^{\theta_{1}-\theta_{2}}
g\left(\frac{1+\alpha_{1}^{2}+\alpha_{2}^{2}}{\alpha_{1}
\alpha_{2}},\theta_{1},\theta_{2}\right),
\end{eqnarray}
where $g$ is an arbitrary function. So the two-point function $G$
is found up to an unknown function:
\begin{eqnarray}
\langle \Phi(z_{1},\bar{z}_{1},\theta_{1})\Phi(z_{2},\bar{z}_{2},
\theta_{2})\rangle&=&
\frac{1}{(y_{1}y_{2})^{2\Delta+\theta_{1}+\theta_{2}}}
\left(\frac{y_{1}}{y_{2}}\right)^{\theta_{1}-\theta_{2}}
h\left(\frac{x^{2}+y_{1}^{2}+y_{2}^{2}}{y_{1}y_{2}}\right),
\end{eqnarray}
which is the same as the solution obtained in \cite{MR}.\\
\vspace{10mm}\\ {\large {\bf Acknowledgement}} We are grateful to
M. Flohr for thorough reading of the manuscript and helpful
comments. We would like also to thank I. I. Kogan, and M. R.
Rahimi-Tabar for discussions and comments.
\section{Appendix A: Level 3 singular vector}
In this appendix we derive level 3 singular vectors. According to
equation (28) the general form of a singular vector at level 3 is:
\begin{eqnarray}
|\chi_{\Delta,c}^{3}(\theta)\rangle
&=&(b^{(1,1,1)}L^{3}_{-1}+b^{(1,2)}L_{-1}L_{-2}+b^{(3)}L_{-3})|\Delta+\theta
\rangle.
\end{eqnarray}
By applying $L_{1}$ on both sides of above equation and use of
Virasoro algebra and equation (18), we have:
\begin{eqnarray}
[6b^{(1,1,1)}(\Delta+1+\theta)L^{2}_{-1}+b^{(1,2)}(3L^{2}_{-1}+2(\Delta+2+\theta)L_{-2})+4b^{(3)}L_{-2}]|\Delta+\theta\rangle=0.
\end{eqnarray}
Because $L_{-1}^{2}$ and $L_{-2}$ are independent we must have
\begin{eqnarray}
6(\Delta+1+\theta)b^{(1,1,1)}+3b^{(1,2)}&=&0 \nonumber\\
2(\Delta+2+\theta)b^{(1,2)}+4b^{(3)}&=&0.
\end{eqnarray}
If we choose $b^{(1,1,1)}=1$ and find $b^{(1,2)}$ and $b^{(3)}$ in
equation (80) we get the singular vector at level 3 for rank 2
Jordan cell as:
\begin{eqnarray}
|\chi_{\Delta,c}^{3}(\theta)\rangle=
\{L^{3}_{-1}-2(\Delta+1+\theta)L_{-1}L_{-2}\nonumber\\
+[(\Delta+1)(\Delta+2)+(2\Delta+3)\theta
] L_{-3}\}|\Delta+\theta\rangle.
\end{eqnarray}
Now if we demand that $L_{2}|\chi^{3}_{\Delta,c}(\theta)\rangle=0$
we have:
\begin{eqnarray}
[-3\Delta^{2}+7\Delta-2-c(\Delta+1)]+\theta[-6\Delta-c+7]&=&0
\end{eqnarray}
and therefore $\Delta$ and $c$ are restricted to the values:
\begin{eqnarray}
c&=&\frac{3\Delta^{2}-7\Delta+2}{\Delta+1}=-6\Delta+7.
\end{eqnarray}
From these relations we observe that $\Delta=-3$ or 1 which
corresponds to $c=25$ and 1 respectively. For these values, the
singular vector is orthogonal to any other vector. In particular,
$\langle\chi_{\Delta,c}^{3}(\theta)|\chi_{\Delta,c}^{3}(\theta)\rangle=0$.

Now we are ready to find the explicit form of the singular
vectors. By using equation (81) and
\begin{eqnarray}
|\chi_{\Delta,c}^{3}(\theta)\rangle
=|\chi_{\Delta,c}^{3}(0)\rangle+\theta|\chi_{\Delta,c}^{3}(1)\rangle,
\end{eqnarray}
we have:
\begin{eqnarray}
|\chi_{\Delta,c}^{3}(0)\rangle &
=&[L^{3}_{-1}-2(\Delta+1)L_{-1}L_{-2}+(\Delta+1)(\Delta+2)L_{-3}]
|\phi \rangle
\end{eqnarray}
and
\begin{eqnarray}
|\chi_{\Delta,c}^{3}(1)\rangle=[L^{3}_{-1}-2(\Delta+1)L_{-1}L_{-2}
+(\Delta+1)(\Delta+2)L_{-3}]|\psi \rangle\nonumber\\
+[-2L_{-1}L_{-2}+(2\Delta+3)L_{-3}]|\phi \rangle.
\end{eqnarray}
As expected equation (85) is exactly the level 3 singular vector
of a normal CFT. But in such theories there is no restriction on
the values of $c$ or $\Delta$. On the other hand in LCFT's the
presence of a second null vector forces us to allow only certain
values of $c$ and $\Delta$ as given by equation (83). We observe
that some null vectors obtained in \cite{Fl-Sing} are missing, in
other words we have an incomplete set. The reason for this may be
that equation (27) sets too strong a condition within LCFT
\cite{Private}.
\section{Appendix B: Hypergeometric Functions}
In this appendix we show that equation (48) can be considered as
16 differential equations. However one of them is trivial because
it vanishes due to OPE constraints \cite{MRS,FlohrNew}. According
to equation (49) $a$, $b$ and $c$, $H$ are functions of
$\theta_{i}$'s. We write any of them in a general form:
\begin{equation}
H=\sum_{i=1}^{4}H_{i}\theta_{i}+\sum_{1\leq
i<j\leq4}H_{ij}\theta_{i}\theta_{j}+ \sum_{1\leq
i<j<k\leq4}H_{ijk}\theta_{i}\theta_{j}\theta_{k}+H_{1234}
\theta_{1}\theta_{2}\theta_{3}\theta_{4}
\end{equation}
and in a similar way for $a$, $b$ and $c$. Now by substitution of
them in equation (48) we obtain 15 differential equations:

\begin{equation}
DH_{i}=0
\end{equation}
\begin{eqnarray}
DH_{ij}=\{-[c_{i}-(a_{i}+b_{i})\eta]
\frac{dH_{j}}{d\eta}+(a_{0}b_{i}+a_{i}b_{0})H_{j}+
i\longleftrightarrow j\}\nonumber\\
\end{eqnarray}
\begin{eqnarray}
DH_{ijk}=[-[c_{k}-(a_{k}+b_{k})\eta]\frac{dH_{ij}}{d\eta}+
(a_{0}b_{k}+a_{k}b_{0})H_{ij}\nonumber\\
-[c_{ij}-(a_{ij}+b_{ij})\eta]\frac{dH_{k}}{d\eta}
+(a_{0}b_{ij}+a_{i}b_{j}+a_{ij}b_{0})H_{k}+cyclic\:\:\: terms ]
\end{eqnarray}
\begin{eqnarray}
DH_{1234}=[-[c_{l}-(a_{l}+b_{l})\eta]\frac{dH_{ijk}}{d\eta}+(a_{0}b_{l}+a_{l}b_{0})H_{ijk}\nonumber\\
-[c_{ij}-(a_{ij}+b_{ij})\eta]\frac{dH_{kl}}{d\eta}+(a_{0}b_{ij}+a_{i}b_{j}+a_{ij}b_{0})H_{kl}\nonumber\\
-[c_{ijk}-(a_{ijk}+b_{ijk})\eta]\frac{dH_{l}}{d\eta}+(a_{0}b_{ijk}+a_{ijk}b_{0}+a_{k}b_{ij}+a_{ij}b_{k})H_{l}\nonumber\\
+cyclic\:\:\:terms]\nonumber\\
\end{eqnarray}
where
\begin{equation}
D:=\eta(1-\eta)\frac{d^{2}}{d\eta^{2}}+[c_{0}-(a_{0}+b_{0}+1)\eta]
\frac{d}{d\eta}-a_{0}b_{0}.
\end{equation}

Let us now obtain from equation (51), first few terms of 16
functions that are solutions of differential equations, given
above for the special case of $\Delta_{i}=\frac{1}{4}$
\begin{eqnarray}
h_{0}&=& F(2,1,2,\eta)=1+\eta+\eta^{2}+\eta^{3}+\cdots\nonumber\\
h_{1}&=&\:\:\:\frac{1}{2}\eta+\frac{2}{3}\eta^{2}+\frac{3}{4}\eta^{3}+\cdots\:\:,\:\:\:\:\:\:\hspace{4mm}h_{2}=\frac{3}{2}\eta+\frac{7}{3}\eta^{2}+\frac{35}{12}\eta^{3}+\cdots\nonumber\\
h_{3}&=&-\frac{1}{2}\eta-\frac{2}{3}\eta^{2}-\frac{3}{4}\eta^{3}+\cdots\:\:,\:\:\:\:\:\:\hspace{4mm}h_{4}=\frac{1}{2}\eta+\frac{7}{9}\eta^{2}+\frac{35}{36}\eta^{3}+\cdots\nonumber\\
h_{12}&=&-\frac{1}{2}\eta-\frac{1}{18}\eta^{2}+\frac{29}{72}\eta^{3}+\cdots\:\:,\:\:\:\:\:\:h_{13}=\frac{1}{2}\eta+\frac{4}{9}\eta^{2}+\frac{3}{8}\eta^{3}+\cdots\nonumber\\
h_{23}&=&-\frac{2}{3}\eta^{2}-\frac{5}{4}\eta^{3}+\cdots\:\:,\:\:\:\:\:\:\hspace{13mm}
h_{14}=\frac{1}{3}\eta+\frac{2}{3}\eta^{2}+\frac{11}{12}\eta^{3}\nonumber\\
h_{24}&=&\frac{7}{6}\eta+\frac{70}{27}\eta^{2}+\frac{281}{72}\eta^{3}+\cdots\:\:,\:\:\hspace{1mm}h_{34}=-\frac{1}{2}\eta-\frac{49}{54}\eta^{2}-\frac{259}{216}\eta^{3}+\cdots\nonumber\\
h_{123}&=&\frac{7}{9}\eta^{2}+\frac{7}{6}\eta^{3}+\cdots\:\:,\:\:\:\:\:\:\hspace{15mm}h_{124}=-\frac{2}{3}\eta-\frac{41}{162}\eta^{2}+\frac{317}{648}\eta^{3}+\cdots\nonumber\\
h_{134}&=&\frac{2}{3}\eta+\frac{121}{162}\eta^{2}+\frac{455}{648}\eta^{3}+\cdots\:\:,\:\:\:\:h_{234}=-\frac{32}{27}\eta^{2}-\frac{46}{18}\eta^{3}+\cdots\nonumber\\
h_{1234}&=&\frac{137}{81}\eta^{2}+\frac{17}{6}\eta^{3}+\cdots.
\end{eqnarray}
where $Dh_{0}=0$.

\end{document}